\documentclass[reprint,amsmath,amssymb,aip,jcp]{revtex4-2}
\usepackage{graphicx}
\usepackage{dcolumn}
\usepackage{bm}
\newcommand{\nc}{\ensuremath{N_\text{C}}}

\begin{document}
%
%\title{Towards a complete characterization of the thermodynamics of isolated hydrocarbon clusters}
\title{Finite-temperature stability of hydrocarbons: fullerenes versus flakes}
\author{Ariel Francis Perez-Mellor}
\altaffiliation{Present address: Department of Physical Chemistry, University of Geneva,
30 Quai Ernest-Ansermet, 1211 Geneva 4, Switzerland}%
\affiliation{Institut des Sciences Moléculaires d'Orsay (ISMO), CNRS, Université Paris Saclay, 91405 Orsay, France}

\author{Pascal Parneix}%
\affiliation{Institut des Sciences Moléculaires d'Orsay (ISMO), CNRS, Université Paris Saclay, 91405 Orsay, France}%
\author{Florent Calvo}
\affiliation{Univ. Grenoble Alpes, CNRS, LIPhy, 38000 Grenoble, France}
\author{Cyril Falvo}
\email{cyril.falvo@universite-paris-saclay.fr}
\affiliation{Institut des Sciences Moléculaires d'Orsay (ISMO), CNRS, Université Paris Saclay, 91405 Orsay, France}
\affiliation{Univ. Grenoble Alpes, CNRS, LIPhy, 38000 Grenoble, France}
\begin{abstract}
The effects of a finite temperature on the equilibrium structures of hydrocarbon molecules are computationally explored as a function of size and relative chemical composition in hydrogen and carbon. Using parallel tempering Monte Carlo simulations employing a reactive force field, we find that in addition to the phases already known for pure carbon, namely cages, flakes, rings
and branched structures, strong effects due to temperature and the
addition of little amounts of hydrogen are reported. Both entropy and
the addition of moderate amounts of hydrogen favor planar structures
such as nanoribbons over fullerenes. Accurate phase diagrams are
proposed, highlighting the possible presence of multiple phase changes
at finite size and composition. Astrophysical implications are also
discussed.
\end{abstract}
\maketitle
%
%%%%%%%%%%%%%%%%%%%%%%%%%%%%%%%%%%%% 
%
\section{Introduction}
Carbon systems from small clusters to bulk matter exhibit a large
variety of structures and properties due to the ability of carbon to
form single, double and triple bonds. In the solid state, carbon is
present as graphite and diamond, but can also be amorphous or even porous,
making it useful for separation or energy conversion purposes.\cite{Bandosz:2020wg} Carbon also
drives a huge interest as the key to advanced low-dimensional materials such as
nanotubes or graphene. Despite being the focus of extensive research
for many
decades,\cite{Helden:1991aa,Handschuh:1995aa,Van-Orden:1998aa,Koyasu:2012ul,Moriyama:2018vg}
many properties of carbon systems remain poorly understood. This is
particularly true for carbon systems under extreme conditions such as
clusters in the gas phase\cite{Van-Orden:1998aa} or liquid
carbon.\cite{Hull:2020to}\par
In the dilute limit, carbon clusters and hydrocarbon molecules play an
important role in astrochemistry. For example, the presence of
polycyclic aromatic hydrocarbons (PAHs) in the interstellar medium
(ISM) was suggested nearly forty years ago through the
observation of the so-called aromatic infrared bands
(AIB).\cite{Leger:1984fk,Allamandola:1985uq,Tielens:2008fk} In
addition, a variety of pure carbon clusters have now been conclusively
observed in the ISM, ranging from small carbon
chains\cite{Bernath:1989uk,Maier:2001tr} up to C$_{60}$ and C$_{70}$
fullerenes.\cite{Cami:2010aa,Sellgren:2010uq} However, a much larger
variety of hydrocarbon systems is expected to be present
under astrophysical conditions, and contribute to different spectral
features of the ISM, such as the diffuse interstellar bands (DIB) and the
ultraviolet (UV) bump, as well as the AIBs
themselves.\cite{Hobbs:2008vt, Steglich:2011er,Gavilan:2017tx,Bonnin:2019aa,Dubosq:2019aa,Dubosq:2020aa,Omont:2021tk,McCarthy:2021vq,Rademacher:2022uf}\par
The allotropy of carbon conveys to the nanoscale, as pure carbon
clusters display a large variety of structures and can be found as
chains, rings, flakes (mostly planar, aromatic structures) and cages
(which include
fullerenes).\cite{Van-Orden:1998aa,Helden:1991aa,Gotts:1995tw,Handschuh:1995aa}
Quite a number of theoretical studies using different potential energy
surfaces have focused on low-energy structures, typically explored as
a function of the number of carbon and hydrogen
atoms.\cite{Taylor:1995wm,Zhang:2002wt,Cai:2004aa,Shao:2006aa,Shao:2007aa,Kosimov:2008wl,Kosimov:2010aa,Yen:2015tt,Lai:2016aa,Varandas:2018wl,Clare:2003tn,Bihlmeier:2008vc,Tachikawa:2014uh,Lepeshkin:2022ua}
While these studies generally highlighted the dependence of the
results on the specific energy model, they all concluded that for
small pure carbon clusters ($\nc\lesssim 8$) the lowest isomers are
linear chains, followed by rings ($8\lesssim \nc \lesssim 16$), flakes
$(16 \lesssim \nc \lesssim 24)$, and fullerenes ($N \gtrsim 24)$.
Additional hydrogen atoms strongly alter the stability of these
structures.\cite{Lepeshkin:2022ua} In particular, it is expected that fullerenes, which are
the most stable structures for large clusters, become less stable as
hydrogen atoms are added. Conversely, flakes structures (which
includes the PAH family) should become edge stabilized by
hydrogenation much more conveniently than fullerenes.\par
Besides structural investigations, only a few studies were aimed at
including the effects of a finite temperature or excess energy on
isolated hydrocarbon
compounds.\cite{Kim:1994aa,Martin:1996tj,Allouch:2021vb} In their
seminal computational study, Kim and Tom\'anek\cite{Kim:1994aa} showed
using a tight-binding model that fullerenes exhibit multiple phase
transitions upon heating. After first losing their well-defined,
highly symmetric structures forming the low-temperature solid state, a
more floppy phase arises that still consists of cages but with
increasing amounts of topological defects. Upon further increasing the
temperature, the fullerenes undergo dramatic phase changes at 4000 K
into a so-called pretzel phase corresponding to interconnected carbon
rings, and eventually into multiple connected chains, until they
finally dissociate at even higher temperatures. The pretzel phase
transition was suggested as the equivalent of the melting phase
transition in bulk carbon.\cite{Kim:1994aa} While these temperatures are notably high
for laboratory experiments, they can be reached rather easily under
astrophysical conditions, even for small isolated molecules that
undergo photonic excitation in the UV range: a single 10 eV photon
absorbed by a 60-atom molecule is already equivalent to 700~K
heating.\par
To a large extent, the phenomenology of fullerene melting explored by
Kim and Tom\'anek\cite{Kim:1994aa} is consistent with the structural
diversity in carbon clusters recently addressed more extensively through systematic sampling
methods.\cite{Bonnin:2019aa,Allouch:2021vb,Furman:2022ur} However,
while the importance of sp$^2$-dominated cages at low temperature or
energy and sp$^1$-dominated pretzels and chains at higher temperatures
could be confirmed, the phase of flakes, still sp$^2$ dominated but
with a much more open character, has remained relatively
overlooked. Yet flakes and graphene nanoribbons are important motifs
of carbon at the nanoscale, not only for their use as building blocks
for innovative
materials\cite{Dutta:2010vz,Georgakilas:2015vh,Wang:2021vz} but also,
in the astrochemical context, as possible intermediates in the
formation of fullerenes themselves.\cite{Berne:2012uu,Berne:2015ta}
Flakes and nanoribbons can be stabilized by the presence of hydrogen,
acting to protect the unsaturated and peripheral carbon atoms that
are specific to this rather open phase. In view of the abundance of
hydrogen in the Universe, it is also essential to determine its
effects on the thermodynamics of dilute carbon matter, and how the
combined influences of temperature, cluster size, and relative amount
of hydrogen affect the preferred structures. Our goal in this article
is to bridge this gap with our current understanding that is mostly
limited to pure carbon, by means of atomistic modeling that includes
exhaustive sampling methods and a reactive potential relevant for hydrocarbon compounds. In particular, by designing appropriate order
parameters, we are able to delineate the stability conditions of
cages, pretzels and chains, but also of the elusive flakes, through
entire finite size phase diagrams.\par
\section{Theoretical methods}
Carbon clusters are characterized by a highly rugged energy landscape
whose exploration requires efficient computational approaches. In this
work, we use the well-established parallel tempering Monte Carlo (PTMC)
method in the canonical ensemble.\cite{Swendsen:1986aa} To prevent
irreversible dissociation that is likely to occur at high enough
temperatures, we further impose a connectivity criterion to the
structures sampled during the Monte Carlo process, rejecting
disconnected configurations. Here, simple distance criteria were used
to define connectivity, with $d_{\text{CC}}=3$~\AA\ for carbon-carbon
distances and $d_{\text{CH}}=d_{\rm HH}=2$~\AA\ for distances
involving hydrogen atoms. These values were chosen significantly
larger than typical chemical bond distances in such a way that bond
breaking and formation are still allowed, while also avoiding
irreversible dissociation. As a result, at high temperatures, the
presence of dissociated structures is expected, but with fragments
that remain close to each other. This also provides a means of
quantifying the stability of the fully connected structures.\par
Following our earlier work,\cite{Bonnin:2019aa} we use the
second-generation reactive empirical bond order (REBO)
potential\cite{Brenner:2002xy} to describe atomic interactions
between carbon and hydrogen atoms in the clusters. This reactive force
field was originally designed to describe the energetic properties of isolated
hydrocarbon molecules as well as various carbon materials in condensed
phases. In particular, it has been used to describe the lowest-energy
structures of carbon clusters\cite{Lai:2016aa} with good agreement
with respect to other methods that explicitly account for electronic
structure, such as the density-functional-based tight-binding
(DFTB) method.\cite{Yen:2015tt}\par
PTMC simulations were performed for 12 different pure carbon clusters
with sizes in the range of 20--80 atoms. Hydrogenated clusters were
also studied for specific amounts of carbon, namely 28 or 60 carbon
atoms and 0--12 or 0--20 hydrogen atoms, respectively. For all systems,
the PTMC simulations employed a total of 28 or 32 replicas with
temperatures allocated according to a geometric progression,
additional temperatures being inserted around the main melting phase
change to further enhance sampling. The complete numerical details of
the PTMC simulations are provided as Supplementary Information. The
raw data from the PTMC simulations were then processed using a version of the
weighted histogram analysis method\cite{Kumar:1992wa} based on an implementation from Poteau et al.\cite{Poteau:1994aa}  to yield the various
properties of interest as a continuous function of temperature.\par
To interpret the thermodynamical observables and connect their
features to the underlying structural properties of the clusters, and
also to merely assign the phases themselves, a number of order
parameters were employed. Following previous works,
\cite{Blavatska:2010zp,Calvo:2012aa,Bonnin:2019aa} the shape of the
clusters is characterized using three rotationally invariant
parameters derived from the successive moments of the gyration tensor,
namely the gyration radius $R_g$, the asphericity $A_3$ and the
prolateness $S$ of the atomic distribution. Altogether, these
quantities, fully explicited in the Supplementary Information, measure
the geometrical extension, the similarity to a sphere, a disk, or a
chain. In addition to providing insight into the geometrical extension of the
system for a given size, the gyration radius is also naturally expected
to be particularly sensitive to size itself. In two-dimensional
flakes, but also in hollow (single shell) cages, $R_g$ should scale
with the number of carbon atoms \nc\ as $\sqrt{\nc}$. A similar
scaling is expected for branched structures according to polymers
theory.\cite{Fixman:1962aa} To better compare $R_g$ across system
sizes, we thus employ a scaled gyration radius $\widetilde R_g =
R_g/\sqrt{\nc}$. For hydrogen-containing systems, and since our interest lies in the effects of hydrogen on the carbon nanostructure, hydrogen atoms are neglected when calculated the gyration tensor and the same definition is used for $\widetilde R_g$.\par
In addition to their geometrical features, clusters were also
characterized in terms of their chemical bonding. Here we have used
simple coordination parameters as measures of hybridization of the
carbon atoms, defining the fractions $\widetilde N_\alpha =
N_\alpha/\nc$ of such atoms having a fixed coordination number
$\alpha=2$ or 3. We use the definition of the atomic coordination
introduced in the REBO potential itself,\cite{Brenner:2002xy} which
for any carbon atom $i$ yields a number $N_i^{\rm C}$ of nearest
neighbors as
\begin{equation}
N_i^\text{C} = \sum_{j\neq i}^{\nc} f_{\text{CC}}^c\left(r_{ij}\right),
\end{equation}
where $f_{\text{CC}}^c\left(r_{ij}\right)$ is a smooth cutoff function
acting between 1.7~\AA\ and 2.0~\AA\ (see
Ref.~\citenum{Brenner:2002xy}).\par
\begin{figure}
\centering
\includegraphics[width=8.5cm]{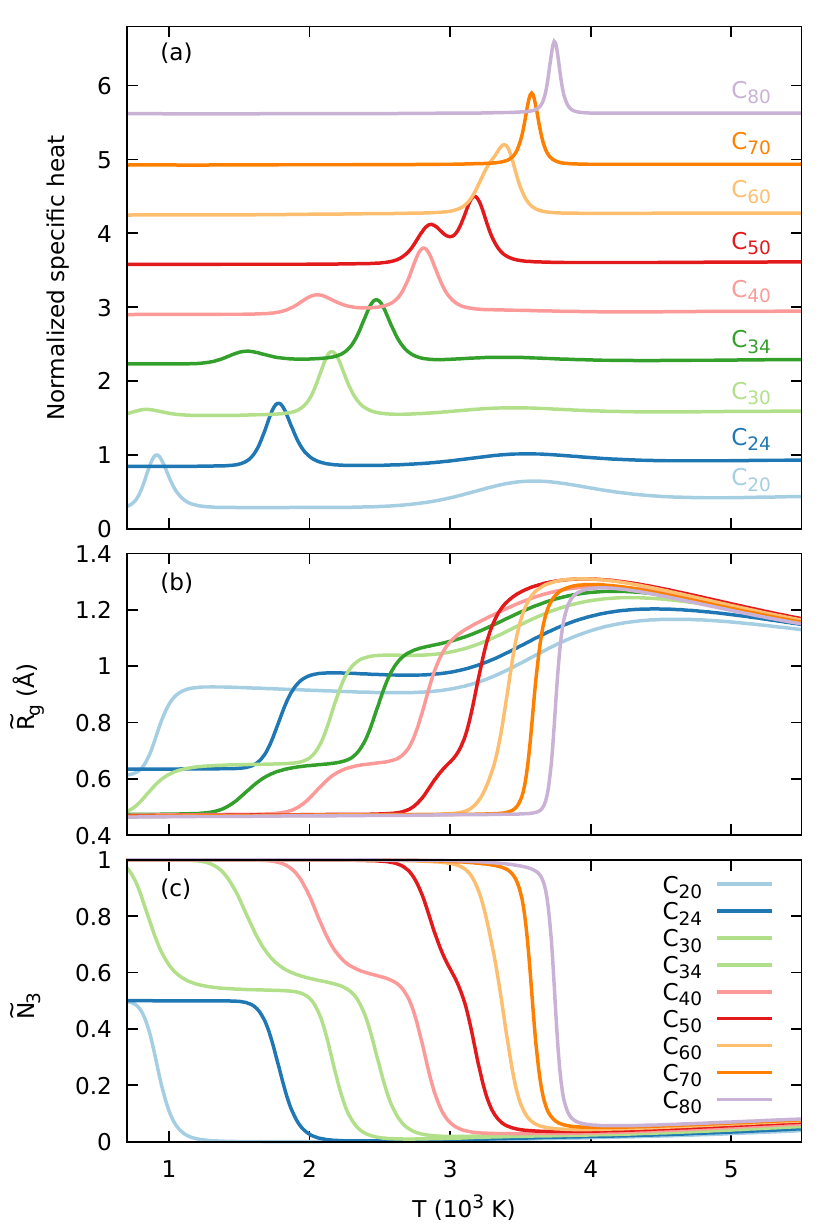}
\caption{Temperature dependence of various properties of pure carbon
  clusters: (a) normalized specific heat at its maximum; (b) average scaled gyration radius;
  (c) average fraction of triply-coordinated carbon atoms. The specific heats curves have been shifted vertically to improve readability.}
\label{fig:carbon_cv_rg_n3}
\end{figure}
\section{Results}
\subsection{Pure carbon clusters}
First-order finite-size phase changes are conveniently identified by sudden
variations in the system energy upon increasing temperature. With
increasing size, these energy variations become sharper until an abrupt
discontinuity (the latent heat) is recovered in the bulk limit of the
proper first-order phase transition. As a result, the specific heat,
which corresponds to the derivative of the internal energy with
respect to temperature, may exhibit one or several prominent peaks at
the corresponding transition temperatures in finite size systems as
well. Fig.~\ref{fig:carbon_cv_rg_n3}(a) shows the specific heat as a
function of temperature for the various carbon clusters, normalized at
their maxima for a better comparison across sizes.\par
For the larger clusters C$_{60}$, C$_{70}$ and C$_{80}$ only a single
peak is found at $T=3390$ K, $T=3580$ K and $T=3740$~K, respectively,
suggesting that these clusters experience only a single phase
change. This phase change corresponds to melting, as already analyzed
in Ref.~\citenum{Kim:1994aa}. In C$_{60}$, the peak appears slightly
asymmetric with a clear shoulder on the low-temperature side. In the
smaller clusters containing 30--50 atoms, an additional peak is found
in the specific heat, which becomes increasingly separated from the
main peak as size decreases. Moreover, the two features shift to lower
temperatures as size decreases. In C$_{34}$ and below, a third feature
can be perceived at temperatures higher than the main peak, in the
3000--4000~K range, which furthermore becomes increasingly well
defined as the clusters become smaller.\par
\begin{figure}
\includegraphics[width=8.5cm]{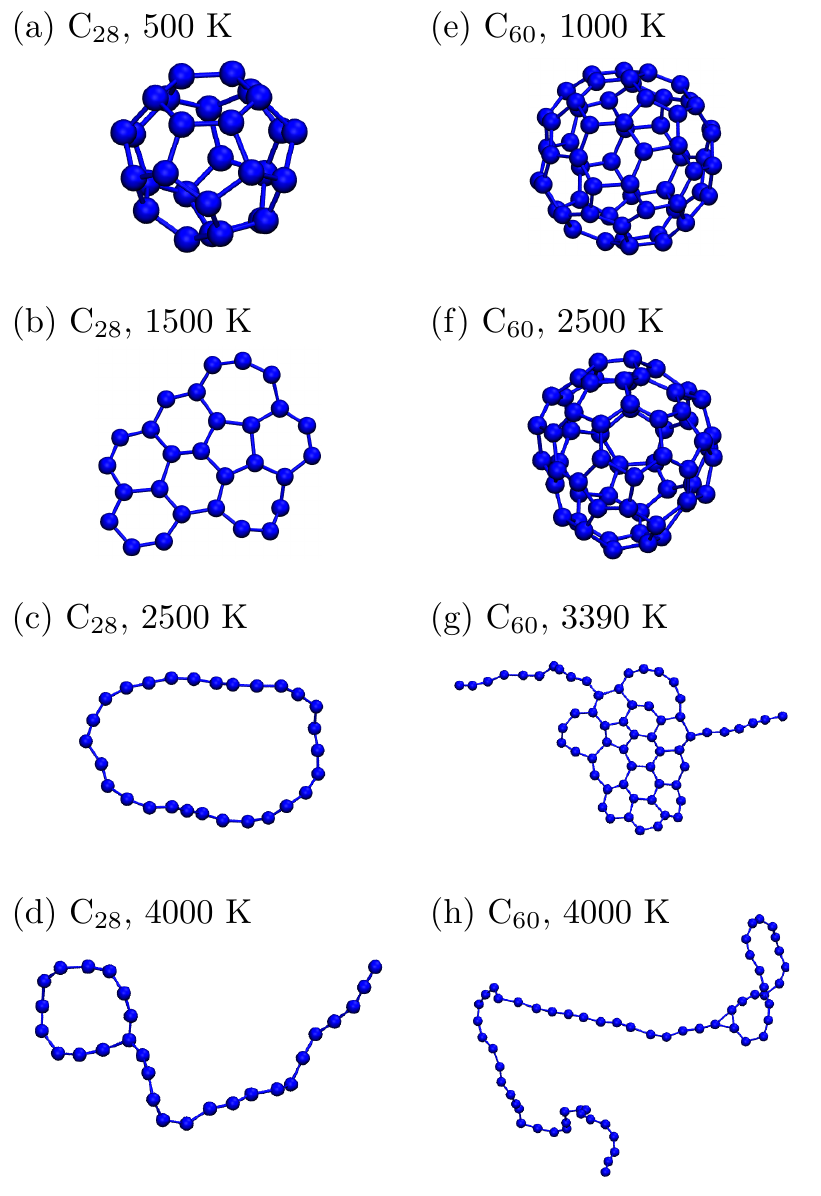}
\caption{Most representative structures for various pure carbon
  clusters at different temperatures, as defined from distances
  based on the minimal set of descriptors.}
\label{fig:carbon_structures}
\end{figure}
To assist in the interpretation of these thermal features,
Figs.~\ref{fig:carbon_cv_rg_n3}(b) and~\ref{fig:carbon_cv_rg_n3}(c)
show the temperature dependence of the average scaled radius of
gyration $\widetilde R_g$ and the proportion $\widetilde N_3$ of
triply-coordinated atoms. The three phase changes identified in the
specific heat are clearly seen with these parameters. For
$\nc\geqslant 28$, the scaled radius of gyration at low temperature is
around $0.47$~\AA\ and all carbon atoms are triply coordinated. This
corresponds to cages and is consistent with the known geometry of
fullerenes. For instance, buckminsterfullerene has a radius of
$3.60$~\AA\ with the REBO potential, corresponding to a value of the
scaled radius of gyration of $0.465$~\AA, a zero-temperature value
that is fully consistent with the data in
Fig.~\ref{fig:carbon_cv_rg_n3}(b).\par
From the PTMC simulations, structures were characterized based on
their values of the different order parameters $R_g$, $A_3$, $S$,
$N_2$ and $N_3$, which form a minimal set of descriptors for the
clusters. At a given temperature, a specific structure can be
retrieved from the sample as the one closest to the configuration with
average values of these descriptors. Here, a simple Euclidean metric
was used to define distances in descriptor space (see Supplementary Information). Several of these
most representative structures are depicted in
Fig.~\ref{fig:carbon_structures} for different relevant temperatures.\par
At low temperature, the most representative structures of C$_{28}$ and
C$_{60}$, as shown in Figs.~\ref{fig:carbon_structures}(a) and
\ref{fig:carbon_structures}(e), are clearly those of cages with little
or no topological defect. As temperature increases and reaches the
3000--4000~K range, clusters with up to 60 carbon atoms lose their
aromaticity with sharp decreases in the numbers of triply-coordinated
atoms and increases in the radii of gyration. The statistical
distributions of these parameters (see Fig.~S2, in Supplementary Information)
remain very narrow around the average at low temperature but, above
the transition temperature, become much broader, particularly for the scaled radius of gyration, showing an increased
structural diversity. The configurational population in this higher
temperature range corresponds to branched and dissociated structures,
see Figs.~\ref{fig:carbon_structures}(d) and
~\ref{fig:carbon_structures}(h). According to our rather loose
criterion for disconnected structures (carbon-carbon distances above
3~\AA), dissociated structures only become significant at temperatures
above the melting range. However, at 4000~K, their proportion remains
rather small, although it steadily increases above this temperature,
reaching 20\% and 47\% at 5500~K for C$_{28}$ and C$_{60}$,
respectively (see Fig.~S3 in Supplementary Information). The continuous
progression between branched and dissociated structures above the
melting temperature indicates that, from a purely thermodynamical
point of view, they can be considered to form a single broad
family.\par
Smaller pure carbon clusters exhibit a different phenomenology, which
can be inferred from the variations in the descriptors.  While
cages are still favored at low temperature, the substantial increase of
$\widetilde R_g$ towards 0.64~\AA\ and the decrease of $\widetilde N_3$
towards 0.6 indicate the predominance of flakes, which is confirmed
by looking at the most representative structures such as the one obtained
for C$_{28}$ at 1500~K in Fig.~\ref{fig:carbon_structures}(b).
Returning to C$_{60}$, a closer inspection of the structural features
indicates that flakes are also found as the main phase but in a rather
narrow temperature range, just below the melting point (see
Fig.~\ref{fig:carbon_structures}(g) for the most representative
structure at 3390~K). This very narrow range of stability, together
with different sampling methods, explain why this phase was not seen
in the earlier work by Kim and Tom\'anek,\cite{Kim:1994aa} especially for fullerenes larger than C$_{60}$. Yet a fourth population of structures
is also visible on the variations of the scaled radius of gyration,
which exhibits values between $\widetilde R_g\approx 0.9$~\AA\ and
$\widetilde R_g\approx1.1$~\AA\ for clusters containing 40 atoms or
less. This population cannot be seen with the triply-coordinated atoms
as they are buried within the family of branched structures. However, a direct
analysis shows that this population corresponds to rings, as
illustrated in Fig.~\ref{fig:carbon_structures}(c) for C$_{28}$ at
2500~K.\par
\subsection{Effect of hydrogenation}
\begin{figure*}
    \centering
    \includegraphics[width=17cm]{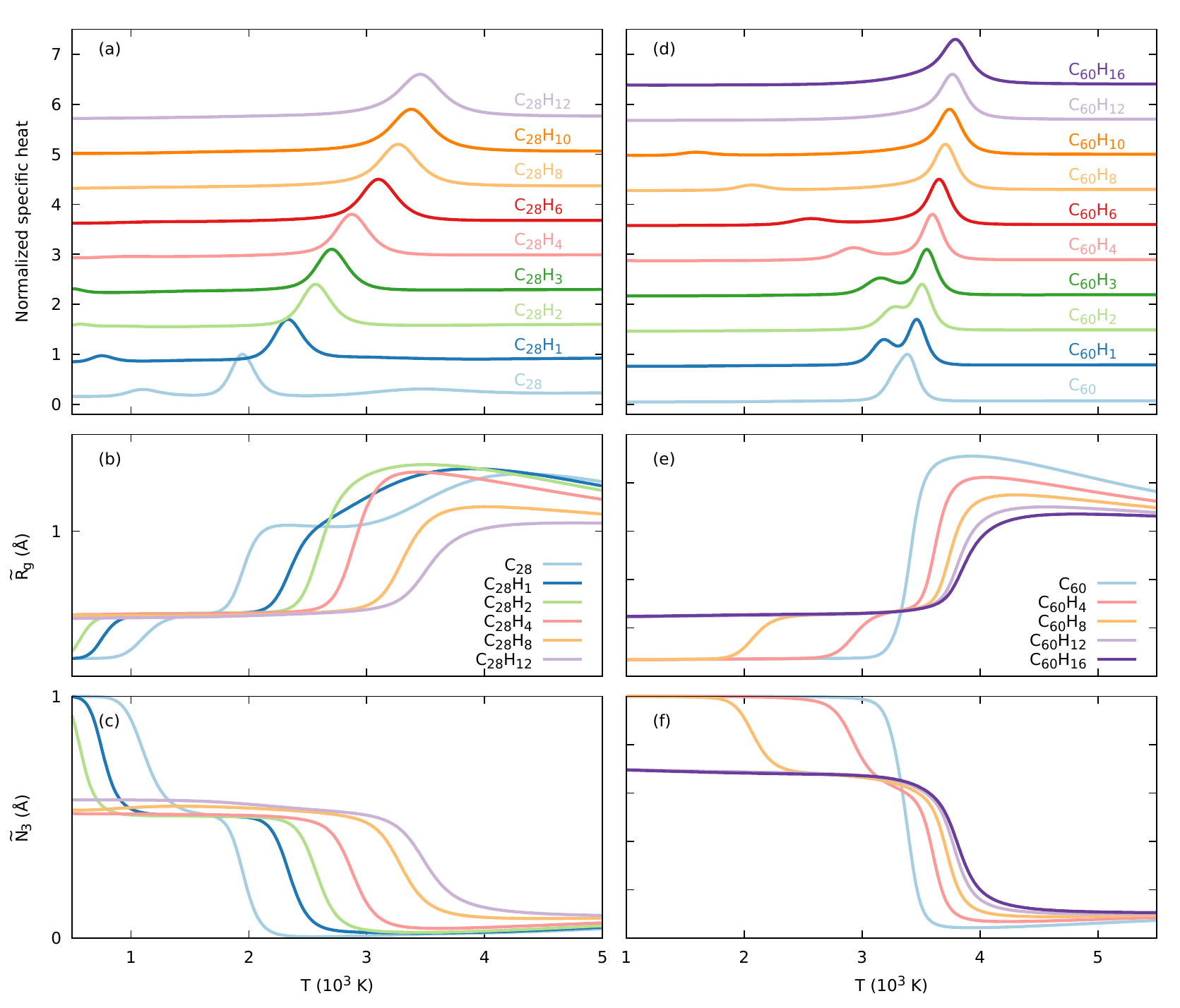}
    \caption{Temperature dependence of various properties of
      hydrocarbon clusters, containing 28 (left panels) or 60 (right
      panels) carbon atoms: (a) and (d) normalized specific heat at its maximum; (b)
      and (e) scaled gyration radius; (c) and (f) fraction of triply-coordinated carbon atoms. The specific heats curves have been shifted vertically to improve readability.}
    \label{fig:hydrocarbon_cv_rg_n3}
\end{figure*}
The above results indicate that the thermodynamics of carbon clusters
depends qualitatively on their size, with up to four possible phases
at finite temperature. We now show that the addition of hydrogen atoms
also strongly affects this thermal behavior.
Fig.~\ref{fig:hydrocarbon_cv_rg_n3}(a) depicts the temperature
dependence of the normalized specific heat for C$_{28}$H$_n$ clusters
with $0\leqslant n\leqslant 12$. As the hydrogen content increases,
the main peak in the specific heat progressively increases from 1960~K
in the pure cluster to 3450~K in C$_{28}$H$_{12}$. Conversely, the
lower temperature peak marking the structural transition from cages to
flakes further shifts to even lower values as more hydrogen is
introduced into the system, becoming no longer discernable above $n=3$.
The broad high-temperature peak found for C$_{28}$ near 3400~K is also
very sensitive to the presence of hydrogen, as it is no longer visible
already for $n=1$.

Strong effects of hydrogen on the phases and phase changes are also
found for C$_{60}$, the specific heats being illustrated in
Fig.~\ref{fig:hydrocarbon_cv_rg_n3}(d) for hydrogen atoms increasing
in numbers from 0 to 16. The single asymmetric peak in the specific
heat found for C$_{60}$ smoothly evolves with increasing hydrogen
loadings, the low-temperature shoulder becoming a distinct peak that
shifts from 3390~K towards lower and lower temperature, reaching
1590~K in C$_{60}$H$_{10}$, while the main melting peak slightly
shifts from 3390~K in the pure carbon cluster to 3780~K in
C$_{60}$H$_{16}$.\par
\begin{figure}
\includegraphics[width=8.5cm]{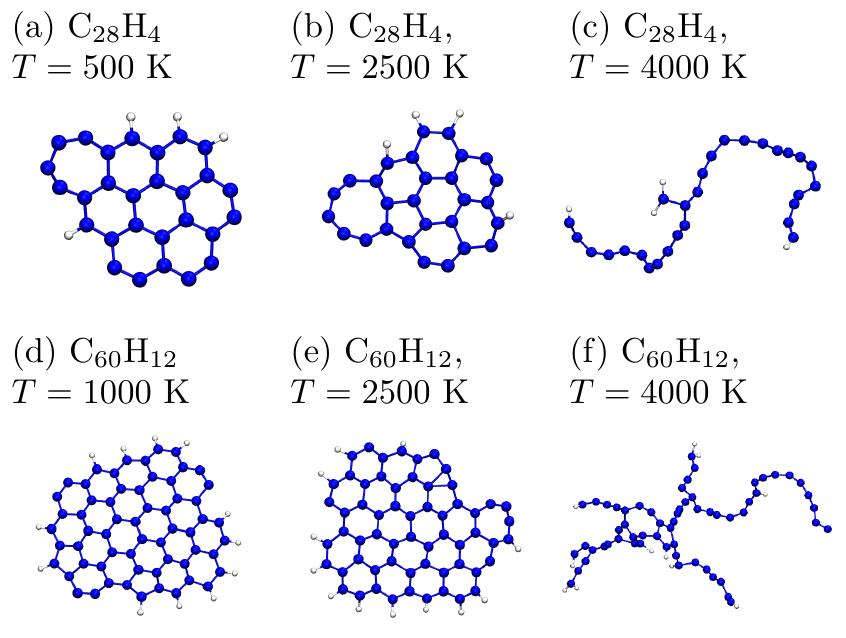}
\caption{Most representative structures for selected hydrocarbon
  clusters at different temperatures, as defined from distances
  based on the minimal set of descriptors.}
\label{fig:hydrocarbon_structures}
\end{figure}
The structural features associated with the clusters underlying the
thermodynamical phases are again inferred by considering the scaled radius of
gyration and the fraction of triply-coordinated carbon atoms, whose
variations with temperature are shown in
Figs.~\ref{fig:hydrocarbon_cv_rg_n3}(b),
\ref{fig:hydrocarbon_cv_rg_n3}(c), \ref{fig:hydrocarbon_cv_rg_n3}(e)
and \ref{fig:hydrocarbon_cv_rg_n3}(f), respectively.

Upon inspection of these quantities, both the cage and ring structures
for C$_{28}$H$_n$ are found to be strongly destabilized by the
addition of hydrogen, and for $n\geqslant4$ only the flake and
branched/dissociated structures remain, as illustrated on their most
representative structures in Fig.~\ref{fig:hydrocarbon_structures}. In
C$_{60}$H$_n$, increasing the hydrogen loading also leads to the
stabilization of the flakes to the expense of cage structures, which
are even no longer found for $n\geqslant 12$, leaving again flakes and
branched or dissociated structures as the only relevant phases. While
the addition of hydrogen generally simplifies the phenomenology of
melting in hydrocarbon clusters at low temperatures, it also affects
the high-temperature behavior by producing greater amounts of
fragments (see Fig.~S3 in Supplementary Information), also with a greater chemical diversity that notably
includes acetylene.\par
\subsection{Phase diagrams}
The results obtained from our simulations can be further processed and
interpreted in terms of finite temperature phase diagrams. More
precisely, and as supported by the sets of representative structures
depicted in Figs.~\ref{fig:carbon_structures} and
\ref{fig:hydrocarbon_structures}, we distinguish cages, flakes,
branched/dissociated structures, as well as rings. The gyration radius
and fraction of triply-coordinated carbon atoms suffice to assign
structures from the three first families unambiguously (see Fig.~S4 in Supplementary Information), but for the
rings it is the fraction $\widetilde N_2 = N_2/\nc$ of
doubly-coordinated atoms that is needed in the absence of atoms with
any greater coordination:
\begin{eqnarray*}
  {\rm cages} &:& \widetilde R_g < 0.56 \\
  {\rm flakes} &:& \widetilde R_g \geqslant 0.56, \widetilde N_3 \geqslant 0.32 \\
  {\rm branched/dissociated} &:& \widetilde R_g \geqslant 0.56, \widetilde N_3  < 0.32 \\
  {\rm rings} &:& \widetilde N_2  = 1
\end{eqnarray*}

\begin{figure*}
    \centering
    \includegraphics[width=17cm]{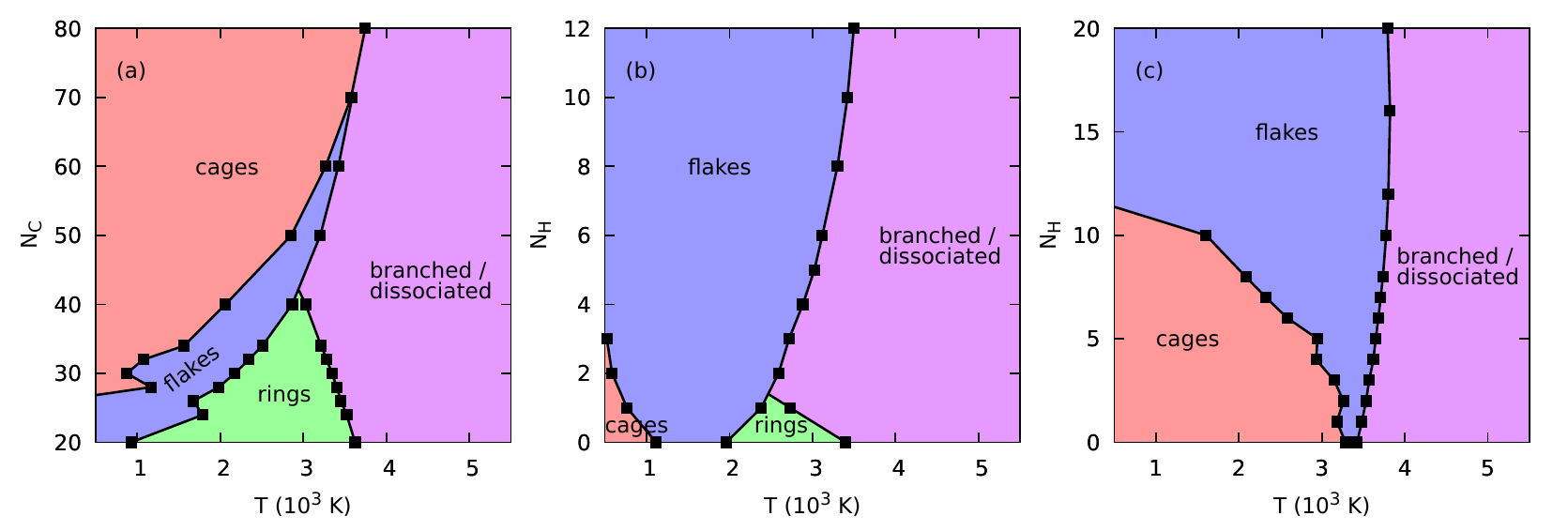}
    \caption{(Temperature, size) stability diagram of (a) pure carbon     
      clusters, (b) C$_{28}$H$_n$; (c) C$_{60}$H$_n$ hydrocarbon clusters.}
    \label{fig:phase}
\end{figure*}

The phase diagrams resulting from these definitions are represented in
Fig.~\ref{fig:phase} as a double function of temperature and numbers of
carbon or hydrogen atoms. Apart from minor size effects,
the boundaries between the various phases are relatively smooth
throughout the entire phase diagrams, suggesting that the trends
inferred from the present simulations are robust.

The structural competition between the four families is usually the
highest in the small size limit, rings being for example stabilized by
entropy ($T>1000$~K) but only for hydrocarbon clusters having no more than 40~atoms. However, the most striking feature of these
diagrams is the competition between cages and flakes, the latter being
favored by hydrogen loading but disfavored by increasing number of
carbon atoms. In particular, our results show that, for the present
model, flakes and graphene nanoribbons are found to be nothing but
metastable configurations for clusters containing more than 70 carbon
atoms, and that they should transform into either cages or branched
structures depending on temperature.

Deeper insight into the specific phase changes can also be obtained by
considering the (Landau) free energy and its variations with
appropriate order parameters. As an example, we provide as
Supplementary Information (Fig. S5) the 1D Landau free energies of
C$_{60}$H$_n$ with $n=0$--10 associated with the scaled radius of
gyration and the proportion of triply-coordinated carbon atoms, at
various temperatures near the phase change between cages and flakes.
Near its melting temperature of 3390~K, C$_{60}$ exhibits two free
energy minima of comparable depth, confirming that the corresponding
transition is first-order-like, rounded by size effects. Another more extended minimum can also be found for the branched and
dissociated structures at large gyration radii or low $N_3$ numbers.
The energy barrier separating cages and flakes from each other is
about 1.7~eV, somewhat lower than Stones-Wales rearrangement energies,\cite{Zhou:2003wc} and remains rather stable
upon moderate hydrogenation for up to $n=8$ hydrogen atoms.
\section{Conclusions}
To summarize, the present computational study provided a fairly
complete statistical survey of the structures of hydrocarbon clusters
at various sizes, compositions, and temperatures in the broad range of
500--5500~K covering the solid-liquid-vapor phases. In large pure
carbon clusters, the results from Kim and Tom\'anek \cite{Kim:1994aa}
were recovered, with melting proceeding through the main phase change
from cages to branched and dissociated structures. However, far more
details could be unravelled here by scrutinizing descriptors aimed at
probing separately the shape and the extent of sp$^2$
hybridization. In particular, the importance of graphene flakes, which
was overlooked in earlier studies, was confirmed and found to have a
thermodynamical signature also in C$_{60}$ as a melting precursor.

In smaller clusters, but also in hydrogenated compounds, the flake
structures become increasingly stable to the expense of cages and the
benefit of (possibly unsaturated) polycyclic aromatic hydrocarbons,
already with few hydrogen atoms. Ring structures also appear to be
stabilized by entropy, contributing to rather rich phase diagrams
exhibiting up to three possible phase changes driven by temperature,
at fixed size and composition.  The structural complexity is
consistent with earlier ion mobility experiments,
\cite{Gotts:1995tw,Lee:1997wj} and our results provide the typical
ranges of excitation energies needed to access the various specific
structures.

Beyond laboratory experiments, our results also have important
implications in astrochemistry, where organic compounds can be heated
quite significantly by absorption of single photons in the UV/visible
range. This in itself would suffice to enable isomerization into a
large diversity of structures, especially graphene flakes, thus
contributing to different absorption and emission bands, and more
generally to spectroscopic features that are much broader than the
highly resolved and so specific fullerenes signatures.

Another important result of the present work is the relatively easy
stabilization of polycyclic aromatic hydrocarbons upon hydrogenation
of pure carbon clusters, even by relatively small amounts of hydrogen.
The connection between fullerene structures and planar flakes, which
has been the subject of earlier works,\cite{Berne:2012uu,Berne:2015ta}
is found here to be further favored by the astrophysical environment.

The present simulations at finite temperature extend the very recent survey of stable hydrocarbon structures at zero temperature (Ref. \citenum{Lepeshkin:2022ua}), and emphasize the key role of entropy in medium size clusters. However, kinetic considerations were ignored and we expect the observation time scale to play an important role on the possible
characterization of the various species identified in the phase
diagrams. The branched/dissociated family is particularly prone to
kinetic effects, as cluster fragmentation could be irreversible
depending on the experimental conditions of density and temperature.
Larger scale rearrangements between cages and flakes, in the context
of fullerene formation,\cite{Berne:2012uu,Berne:2015ta}
would also be worth scrutinizing further,
possibly through transition path sampling
approaches\cite{Dellago:1998dz,Cabriolu:2017aa} but also using tools from graph theory.\cite{Perez-Mellor:2021vw}

In the astrochemical context, the internal energy of individual
clusters could be transferred to other degrees of freedom through
various photophysical processes, such as internal conversion but also
radiation in the infrared or even optical range through the so-called
Poincaré
fluorescence.\cite{Leger:1988th,Martin:2013wt,Lacinbala:2022uj} In
future contributions, it will be important to account for the various
competing kinetic processes and extend the present findings to the
out-of-equilibrium situation of isomerizing clusters and assessing the
importance of radiative cooling.

\section*{Supplementary Material}
A detailed description of the computational methods and structural
parameters; Temperature evolution of the distribution of structural
parameters as well as dissociation probability; Landau free energies
along selected order parameters.

\begin{acknowledgements}
Financial support by the ANR ``PACHYNO'' ANR-16-CE29-0025 is
gratefully acknowledged.  The authors would also like to acknowledge
the computational resources provided by the GRICAD infrastructure
(https://gricad.univ-grenoble-alpes.fr), which is supported by
Grenoble research communities.
\end{acknowledgements}

\section*{Data Availability Statement}

The data that support the findings of this study are available from the corresponding author upon reasonable request.

\bibliography{biblio}% 

\end{document}